\documentclass[aps,showpacs,prep,graphics,twocolumn]{revtex4}
\usepackage{amssymb}
\usepackage[dvips]{graphicx}
\usepackage{amsmath}
\usepackage{bm}
\usepackage{epsfig}

\begin{document}

\title{Present accelerated expansion of the universe from new Weyl-Integrable gravity approach}

\author{$^{1}$ Ricardo Aguila, $^{1}$ Jos\'e Edgar Madriz Aguilar \thanks{E-mail address: madriz@mdp.edu.ar}, $^{1}$ Claudia Moreno  and $^{2,3}$ Mauricio Bellini
\thanks{E-mail address: mbellini@mdp.edu.ar} }
\affiliation{$^{1}$ Departamento de Matem\'aticas, Centro Universitario de Ciencias Exactas e ingenier\'{i}as (CUCEI),
Universidad de Guadalajara (UdG), Av. Revoluci\'on 1500 S.R. 44430, Guadalajara, Jalisco, M\'exico.  \\
E-mail: edgar.madriz@red.cucei.udg.mx\\
and \\
$^{2}$ Departamento de F\'isica, Facultad de Ciencias Exactas y Naturales, Universidad Nacional de Mar del Plata (UNMdP),
Funes 3350, C.P. 7600, Mar del Plata, Argentina \\
E-mail: mbellini@mdp.edu.ar\\
$^{3}$ Instituto de Investigaciones F\'{i}sicas de Mar del Plata
(IFIMAR)- Consejo Nacional de Investigaciones Cient\'{i}ficas y
T\'ecnicas (CONICET) Argentina.}

\begin{abstract}
We investigate if a recently introduced formulation of general relativity on a Weyl-integrable geometry, contains cosmological solutions exhibiting acceleration in the present cosmic expansion. We derive the general conditions to have acceleration in the expansion of the universe and obtain a particular solution for the Weyl scalar field describing a cosmological model for the present time in concordance with the data combination Planck + WP + BAO + SN.
\end{abstract}

\maketitle

\vskip .5cm
Weyl-Integrable geometry, FRW metric, conformal invariance, cosmological solutions, present cosmic acce\-le\-ra\-ted expansion.

\section{Introduction}

The present accelerated expansion of the universe has become an interesting topic, not just for the absence of a fully satisfactory explanation of its origin, but for the wide range of gravity theories introduced in the quest for viable answers to the problem. Many have been the attempts to construct viable models. This issue has been addressed 
basically in two approaches. In one of them the acceleration is generated by an extra material component in the universe, the dark energy (in which is included the cosmological constant). In the other the acceleration is a purely gravitational effect within the framework of gravitation theories alternative to Einstein's general relativity. Modifications to Einstein's gravitational theory go back to the work of Eddington and Schr\"odinger \cite{Goenner}. Physical scalar fields have been included  in some dark energy models, but unfortunately not always with a solid motivation for its introduction beyond to justify the acceleration in the expansion of the universe. However, none of these is free from problems, as for some quintessence fields \cite{Quint}. The majority of quintessence models have been proposed in the light of general relativity. However, in the second approach, the Weyl-integrable geometry provides a solid tool to incorporate a scalar field as a part of the geometry of the spacetime. A consequence of this is that the scalar field can also describe the gravitational field (in addition to the metric tensor $g_{\mu\nu}$), in a new version of scalar-tensor theory \cite{Carlos1,Carlos2}.\\

The Brans-Dicke (BD) theory of gravity is the simplest scalar-tensor theory considered as an alternative to general relativity, in which gravity is described by both a tensor field $g_{\mu\nu}$  and a scalar field $\Phi$ . In this theory the scalar field, which is not of geometric nature, is not a matter field, instead it determines the inverse of the gravitational coupling parameter and in this sense it is part of gravity. This is the reason why the BD scalar field does not appear in the geodesic equation for both massive and massless particles \cite{BDcite1,BDcite2}. This aspect may be considered as inconvenient in the sense that even when $\Phi$ is a part of the gravitational field, it does not appear as a part of the geometry, which in this case is the Riemannian one.\\

A more congruent gravitational theory of the BD type would be one in which the BD scalar field would play an active role in the dynamical field equations of the theory, the same as in its geometrical structure, describing together with the metric tensor $g_{\mu\nu}$, the gravitational field. Recently Romero and co\-lla\-bo\-ra\-tors, by means of the Palatini variational method, found that the scalar field of a BD theory of gravity can turn the space time geometry (assumed Riemannian in DB theories) into a Weyl-integrable one, generating in this manner a scalar-tensor theory that differs from the ori\-gi\-nal BD one \cite{Romero1}. They also found that the Weyl-Integrable geometry contains a Riemannian structure which allows one to formulate general relativity on  Weyl-integrable manifolds \cite{Carlos2}.\\

A similar result has been obtained in \cite{Quiros}. In this letter a conformal equivalence principle is postulated which mathematically is associated with the conformal invariance of the field equations of the theory of gravitation considered. The conformal equivalence principle is formulated as follows: {\it the laws of gravity look the same, no matter which one of the different conformally related frames is chosen to describe them}. Under this point of view is analyzed an alternative interpretation of conformal transformations of the metric, establishing that they can be viewed as a mapping between Riemann and Weyl-integrable geometries. Thus, when the conformal equivalence principle is assumed to be valid, the transformations relate complementary geometrical pictures of the same physical reality. As an example, it is shown that in order to have a BD theory of gravity in concordance with the conformal equivalence principle, the background geometry of the BD theory must be the Weyl-integrable geo\-me\-try. \\

Thus, in view of the preview results, the Weyl-integrable geometry as gravity theory has several inte\-res\-ting aspects by which to address the problem of the acceleration in the cosmic expansion. In this letter we investigate if the formulation of the general theory of relativity on Weyl-integrable geometry, contains cosmological solutions compatible with an accelerated expansion of the universe, without the introduction of a dark energy component. For this purpose the letter is organized as follows. In Sect. II we give a review of the proposal of C. Romero and collaborators to formulate a general theory of relativity on a Weyl-integrable geometry \cite{Carlos2}. In Sect. III we establish the dynamical field equations of the gravity model on cosmological scales. In Sect. IV we derive the general conditions under which the cosmological solutions of the theory exhibit accelerated expansion on cosmological scales, and we find a particular solution compatible with the data combination Planck+WP+BAO+SN \cite{data}. Finally, in Sect. V we develop some final comments.

\section{A new approach of Weyl-integrable geometry}

As is well-known the Weyl geometry is the simplest extension of Riemann geometry. The general theory of relativity is formulated on the base of Riemannian geo\-me\-try. Unlike Riemannian geometry, in Weyl geometry the nonmetricity condition has a different form.  In a coordinate chart the Weyl nonmetricity condition reads \cite{Quiros,Carlos2}
\begin{equation}\label{w1}
^{(w)}\nabla_{\alpha}g_{\mu\nu}=-\sigma_{\alpha}g_{\mu\nu},
\end{equation}
where $^{(w)}\nabla_{\alpha}$ is the Weyl covariant derivative, $\sigma_{\alpha}$ is a 1-form field also known as gauge vector field, and $g_{\mu\nu}$ are the covariant components of the tensor metric. It can easily be shown that the nonmetricity condition (\ref{w1}) is invariant under the Weyl rescaling transformations,
\begin{equation}\label{w2}
\bar{g}_{\alpha\beta}=\Omega^2g_{\alpha\beta},\quad \bar{\sigma}_{\mu}=\sigma_{\mu}-2\partial_{\mu}ln\Omega,
\end{equation}
being $\Omega(x)$ a non-vanishing differentiable function. In Weyl geometry the affine connection is assumed torsionless and hence the condition (\ref{w1}) yields
\begin{equation}\label{w3}
\Gamma^{\alpha}_{\mu\nu}=\left\lbrace _\mu ^{\alpha}\, _\nu\right\rbrace +\frac{1}{2}g^{\alpha\beta}\left[g_{\beta\mu}\sigma_{\nu}+g_{\beta\nu}\sigma_{\mu}-g_{\mu\nu}\sigma_{\beta}\right],
\end{equation}
where $\left\lbrace _\mu ^{\alpha}\, _\nu\right\rbrace$ denotes the Christoffel symbols.\\

The presence of $\sigma_{\mu}$ in (\ref{w1}) has its consequences when parallel transport of vectors is implemented. One of them is that the length of a vector, $l^2=g_{\alpha\beta}l^{\alpha}l^{\beta}$ varies from point to point even on a closed path: $l=l_0 exp \oint dx^{\mu}\sigma_{\mu}/2$. This effect is known as `` the second clock effect'' which basically consists in the broadening of the atomic spectral lines of the electrons immersed in the $\sigma_{\alpha}$ field. This second clock effect is unobserved, as it was pointed out by Einstein, and thus the Weyl geometry was considered not viable \cite{Cheng}.\\

Subsequently, Weyl proposed a particular subclass of his geometry known as Weyl-integrable (WI) geometry, which does not suffer from the second clock effect. This achievement was due to the fact that Weyl expressed the 1-form field as the gradient of a scalar field: $\sigma_{\mu}=\partial_{\mu}\varphi$. This scalar field is geometrical in nature and is known as the Weyl scalar field. In WI geometry the nonmetricity condition (\ref{w1}) reads
\begin{equation}\label{w4}
^{(w)}\nabla_{\alpha}g_{\mu\nu}=-\partial_{\alpha}\varphi \, g_{\mu\nu},
\end{equation}
while the torsionless affine connection (\ref{w3}) leads to
\begin{equation}\label{w5}
\Gamma^{\alpha}_{\mu\nu}=\left\lbrace _\mu ^{\alpha}\, _\nu\right\rbrace +\frac{1}{2}g^{\alpha\beta}\left[g_{\beta\mu}\partial_{\nu}\varphi+g_{\beta\nu}\partial_{\mu}\varphi-g_{\mu\nu}\partial_{\beta}\varphi\right],
\end{equation}
which is the WI connection.
Along a closed path the Stokes theorem ensures that
\begin{equation}\label{w6}
\oint \frac{\sigma_{\mu}}{2}dx^{\mu}=\oint \frac{\partial_{\mu}\varphi}{2}dx^{\mu}=\oint \frac{d\varphi}{2}=0,
\end{equation}
and thus the length of a vector is preserved when it is parallel transported  along a closed path. However, in the case of an open path the length of a vector continues varying from
point to point. This variation is due to the variation of the scalar product of two vectors $g(u(\lambda), v(\lambda))$ when they are parallel transported along a path $C$ characterized by the parameter $\lambda$. If $\lambda_0$ corresponds to a point $a$ and $\lambda$ corresponds to a different point $b$, the scalar product is given by
\begin{equation}\label{w7}
g(v(\lambda),u(\lambda))=g(v(\lambda_0),u(\lambda_0))e^{-[\varphi(x(\lambda))-\varphi(x(\lambda_0))]}.
\end{equation}
This expression is interpreted as the Weyl scalar field, is responsible for the non-invariance of the scalar product along an opened path, and this is the interpretation usually found in the literature. \\

However, recently Romero and collaborators have shown that we can have a novel interpretation  of (\ref{w7}), in which a Riemannian structure can be recovered into the Weyl-integrable geometry. To show it, they rewrite the equation (\ref{w7}) in the form
\begin{equation}\label{w8}
e^{\varphi(x(\lambda))}g(v(\lambda),u(\lambda))=e^{\varphi(x(\lambda_0))}g(v(\lambda_0),u(\lambda_0)).
\end{equation}
This equation can be interpreted as there being an isometry between the tangent spaces of the manifold (spacetime) at the points $a=C(\lambda_0)$ and $b=C(\lambda)$ only in the effective metric $\hat{g}_{\mu\nu}=e^{\varphi}g_{\mu\nu}$. It is easy to see that with this effective  metric the nonmetricity condition (\ref{w4}) becomes
\begin{equation}\label{w9}
\nabla_{\alpha}\hat{g}_{\mu\nu}=0,
\end{equation}
which corresponds to a Riemannian nonmetricity condition. Moreover, as $\hat{g}_{\mu\nu}=e^{\varphi}g_{\mu\nu}$ is invariant under Weyl rescaling transformations, then any geometrical object constructed only with $\hat{g}_{\mu\nu}$ is also invariant. The isometry acts as a correspondence  between a Weyl frame $(M,g,\varphi)$ and a unique Riemannian frame $(M,\hat{g}=e^{\varphi}g,0)$. In this sense the isometry in (\ref{w8}) implies that a new kind of invariance can be established and the same physical phenomena may appear in different representations.  For example, the present accelerated expansion of the universe may be addressed in both frames, however, a possible advantage is that in a Weyl frame we can find a more satisfactory solution to the problem, taking into account that in  WI gravity the gravitational field has a scalar-tensor nature, whereas in the Riemann frame gravity is just described by a tensor field. Thus, as a result it is convenient to study the possibility that the present acceleration in the expansion of the universe could be explained simply as an effect of gravity, without the need of any dark energy. The answer will be investigated in the next sections.\\

\section{The field equations on cosmological scales}

Let us start considering the simplest action for a Weyl-integrable (WI) gravity \cite{Carlos2,Quiros}
\begin{equation}\label{a1}
^{(w)}\!S=\int d^{4}x\sqrt{-g}\,e^{\varphi}\left[\,^{(w)}\! R+\kappa e^{\varphi} {\cal L}_{m}\right]
\end{equation}
$^{(w)}\! R$ being the Ricci scalar calculated with the WI co\-nnec\-tion, $\kappa$ is a coupling constant, and $g$ is  the determinant of the metric $g_{\alpha\beta}$. This action respects the conformal equivalence principle. When a Palatini variational principle is applied to this action, the appropriate background geometry as a result is the Weyl-integrable geometry \cite{Quiros}. The action (\ref{a1}) must be invariant under the Weyl-integrable rescaling transformations $\bar{g}_{\alpha\beta}=e^{f}g_{\alpha\beta}$, $\bar{\varphi}=\varphi-f$, with $f$ being a coordinate dependent smooth function. The Riemann frame can be obtained simply when we make the particular choice $f=\varphi$ in the WI rescaling transformations i.e. with this election we pass from an arbitrary Weyl frame $(M,g,\varphi)$ to the Riemann frame $(M,\bar{g}=e^{\varphi}g,\bar{\varphi}=0)$.  In the Riemann frame the action (\ref{a1}) becomes the usual action for the standard theory of general relativity, where the Weyl scalar field $\bar{\varphi}$ becomes null. The sources of matter are described by the Lagrangian density ${\cal L}_{m}$ which is con\-si\-de\-red independent of the Weyl scalar field. \\
Employing the expression
\begin{equation}
^{(w)}\!R=R-3\Box\varphi - \frac{3}{2}g^{\mu\nu}\varphi_{,\mu}\varphi_{,\nu}
\end{equation}
and avoiding divergence terms, the field equations derived from the action (\ref{a1}) can be written as
\begin{eqnarray}
G_{\mu\nu} &-& \nabla_{\mu}\nabla_{\nu}\varphi +g_{\mu\nu}\Box\varphi +\nonumber \\
\label{a2}
&+&\frac{1}{2}\left(\varphi_{,\mu}\varphi_{,\nu}+\frac{1}{2}g_{\mu\nu}\varphi_{,\sigma}\varphi^{,\sigma}\right)=\kappa e^{\varphi}T_{\mu\nu},\\
\label{a3}
&&\Box\varphi + \frac{1}{2}g^{\mu\nu}\varphi_{,\mu}\varphi_{,\nu}-\frac{R}{3}= \kappa e^{\varphi}T,
\end{eqnarray}
where $T=g^{\mu\nu}T_{\mu\nu}$ is the trace of the energy-momentum tensor, $G_{\mu\nu}=R_{\mu\nu}-(1/2)Rg_{\mu\nu}$ is the Einstein tensor, $\nabla_{\alpha}$ is the covariant derivative calculated with the Christoffel symbols and $\Box=g^{\alpha\beta}\nabla_{\alpha}\nabla_{\beta}$ is the D'Alembertian operator. It is important to stress that in WI gravity the gravitational field is given by the pair $(g_{\mu\nu},\varphi)$ and thus $\varphi$ is a geometrical scalar field that describes gravity, so it is not a matter field.\\

In order to consider some cosmological implications from this theory,  we introduce the Friedmann-Robertson-Walker (FRW) line element
\begin{equation}\label{a4}
dS_4^2=dt^2-a^2(t)dS_3^2,
\end{equation}
being $dS_{3}^2=\delta_{ij}dx^{i}dx^j$ the spatial 3D Minkowskian line element and $a(t)$ is the cosmological scale factor. For observational reasons we will only consider models with 3D spatially flat curvature.\\

Thus, regarding a perfect fluid with total energy density $\rho_{T}$ and a total pressure $p_T$, for the metric in (\ref{a4}) the field Eq. (\ref{a2}) yield
\begin{small}
\begin{eqnarray}\label{a5}
&&3\left(H+\frac{1}{2}\dot{\varphi}\right)^2-\frac{1}{a^2}\nabla^{2}\varphi
-\frac{1}{4a^2}(\nabla\varphi)^2=\kappa e^{\varphi}\rho_T,\\
&& 2\frac{\ddot{a}}{a}+\left(H+\frac{1}{2}\dot{\varphi}\right)^2+
\ddot{\varphi}+3H\dot{\varphi}-\frac{2}{3a^2}\nabla^{2}\varphi-\nonumber \\
\label{a6}
&& - \frac{5}{12a^2}(\nabla\varphi)^2= -\kappa e^{\varphi}p_T,\\
\label{a7}
&&\varphi_{,i,j}-\frac{1}{2}\varphi_{,i}\varphi_{,j}= 0, \quad i\neq j,
\end{eqnarray}
\end{small}
where $H=\dot{a}/a$ is the Hubble parameter and the dot denote the derivative with respect the cosmic time $t$. Similarly, (\ref{a3}) now reads
\begin{eqnarray}
&&\ddot{\varphi}+3H\dot{\varphi}+\frac{1}{2}\dot{\varphi}^2-\frac{1}{a^2}\nabla^2\varphi - \nonumber\\
\label{a8}
&&-\frac{1}{2a^2}(\nabla\varphi)^2-2(\dot{H}+2H^2)=\kappa e^{\varphi}(\rho_T-3p_T),
\end{eqnarray}
which is not an independent equation.\\

Now, in order to implement the cosmological principle on large scales, let us assume that the Weyl scalar field has two contributions, one on cosmological scales and another one on smaller scales. Thus we use the separation formula
\begin{equation}\label{a9}
\varphi(t,x^{i})=\phi(t)+\delta\varphi(t,x^{i}),
\end{equation}
where $\phi(t)$ is the part on cosmological scales that satisfies the cosmological principle and $\delta\varphi(t,x^{i})$ is valid on smaller scales.
Once we assume (\ref{a9}), it as a result is natural to consider also two contributions for the total energy density and the total pressure in the form
\begin{equation}\label{rp1}
 \rho_T(t,x^i)=\rho(t)+\delta\rho(t,x^i),\quad p_T(t,x^i)=p(t)+\delta p(t,x^i),
\end{equation}
where $\rho(t)=\rho_m(t)+\rho_r(t)$ and $p(t)=p_m(t)+p_r(t)$, being $\rho_m$ and $\rho_r$ the energy density for matter (baryonic matter and dark matter) and the energy density for radiation, while $p_m$ and $p_r$ denote the pressure for matter and radiation, respectively. \\

Under these conditions (\ref{a5}) and (\ref{a6}) read on cosmological scales as
\begin{eqnarray}\label{a10}
3\left(H+\frac{1}{2}\dot{\phi}\right)^2 &=& \kappa e^{\phi}\rho,\\
\label{a11}
2\frac{\ddot{a}}{a}+H^2+\ddot{\phi}+4H\dot{\phi}+\frac{1}{4}\dot{\phi}^2 &=& -\kappa e^{\phi}p,
\end{eqnarray}
where we have considered that on cosmological scales the condition $|\delta\varphi|\ll |\phi|$ holds, thus taking account of the fact that on cosmological scales the astrophysical contributions to the Weyl scalar field are subdominant.

On smaller scales (astrophysical scales mainly) the system (\ref{a5})-(\ref{a6}) yields
\begin{small}
\begin{eqnarray}
&&3(H+\frac{1}{2}\dot{\phi})\delta\dot{\varphi}+\frac{1}{4}\delta\dot{\varphi}^2-\frac{1}{a^2}\nabla^2\delta\varphi-\frac{1}{4a^2}(\nabla\delta\varphi)^2 = \nonumber \\
\label{a12}
&&=\kappa e^{\phi}e^{\delta\varphi}\delta\rho,\\
&&\delta\ddot{\varphi}+4H\delta\dot{\varphi}+\frac{1}{4}(2\dot{\phi}\delta\dot{\varphi}+\delta\dot{\varphi}^2)-\frac{2}{3a^2}\nabla^2\delta\varphi-\nonumber \\
\label{a13}
&&-\frac{5}{12a^2}(\nabla\delta\varphi)^2 = -\kappa e^{\phi}e^{\delta\varphi}\delta p.
\end{eqnarray}
\end{small}
Notice that the dynamics of the Weyl scalar field on astrophysical scales is depending of its cosmological solution part. The dynamical Eqs. (\ref{a12})-(\ref{a13}) can be simplified if we assume on astrophysical scales validity of the opposite condition: $|\delta\varphi|\gg |\phi|$, however, this analysis is out of the scope of this letter.

\section{Cosmological  solutions exhibiting  accelerated expansion}

Now we are in position to investigate if there exist solutions of the new Weyl-integrable gravitational approach, capable to describe the present period of accelerated expansion of the universe. In order to do so, let us to derive the general conditions for accelerated expansion solutions.\\

The cosmological dynamics is described by (\ref{a10}) and (\ref{a11}). Hence, after straightforward calculations theese equations yield
\begin{equation}\label{a14}
\frac{\ddot{a}}{a}=-\frac{1}{6}\kappa e^{\phi}\left[(\rho+3p)+\frac{3}{\kappa}\left(\ddot{\phi}+3H\dot{\phi}\right)e^{-\phi}\right].
\end{equation}
It can easily be seen from this equation that to achieve $\ddot{a}>0$ solutions, the condition
\begin{equation}\label{a15}
(\rho+3p)+\frac{3 e^{-\phi}}{\kappa}\left(\ddot{\phi}+3H\dot{\phi}\right)<0,
\end{equation}
must hold. For  the barotropic equations of state for matter and radiation: $p_m=0$, $p_r=1/3\rho_r$, the condition (\ref{a15}) becomes
\begin{equation}\label{a16}
\left(\frac{3H_0^2}{8\pi G}\right)(\Omega_m+\Omega_r)+\frac{3 e^{-\phi}}{\kappa}\left(\ddot{\phi}+3H\dot{\phi}\right)<0,
\end{equation}
where $\Omega_m$ and $\Omega_r$ are the density parameter for matter and radiation, respectively. \\

The inequality (\ref{a16}) can be satisfied if the following conditions are valid:
\begin{eqnarray}\label{a18}
&&\ddot{\phi}+3H\dot{\phi}<0, \\
\label{a19}
&&\left(\frac{3H_0^2}{8\pi G}\right)(\Omega_m+\Omega_r)<\frac{3e^{-\phi}}{\kappa}\left|\ddot{\phi}+3H\dot{\phi}\right|.
\end{eqnarray}
As is usually done in scalar-tensor theories, for a power law expanding universe with $H=p/t$, a power law Weyl scalar field $\phi(t)=\phi_0(t_0/t)^n$ (here $\phi_0$ is a constant) satisfies the condition (\ref{a18}) only when $\lbrace n>0, n<3p-1, p>1/3\rbrace$. Hence the expression (\ref{a19}) leads to
\begin{equation}\label{a20}
\left(\frac{3H_0^2}{8\pi G}\right)(\Omega_m+\Omega_r)<\frac{3}{\kappa t^2}e^{-\phi}\left|n(n+1-3p)\phi\right|.
\end{equation}
The deceleration parameter is in this case
\begin{equation}\label{a21}
q(t)=\frac{1}{2}\left[1+3\frac{\frac{\kappa e^{\phi}}{3}\left(\frac{3H_0^2}{8\pi G}\right)\Omega_r+\left(\frac{n(n+1)}{t^2}-\frac{4nH}{t}\right)\phi+\frac{n^2}{4t^2}\phi^2}{\kappa e^{\phi}\left(\frac{3H_0^2}{8\pi G}\right)(\Omega_m+\Omega_r)-\frac{3nH}{t}\phi+\frac{3n^2}{4t^2}\phi^2}\right],
\end{equation}
which tends asymptotically to $-1$.\\

As we have mentioned the Weyl scalar field is geometrical in origin and cannot be considered as another component of the cosmic fluid. It is in fact part of the gravitational field. However, in order to have a comparison with observational data we can assume, just for this purpose, that the Weyl scalar field describes a scalar field fluid characterized by an energy density $\rho_\phi$ and a pressure $p_\phi$. If it is the case, it follows from the dynamical equations (\ref{a10}) and (\ref{a11}) that
\begin{eqnarray}\label{nva1}
\rho_{\phi} &=& -\frac{3}{\kappa}\left(H\dot{\phi}+\frac{1}{4}\dot{\phi}^2\right)e^{-\phi}\\
\label{nva2}
p_{\phi}&=&\frac{1}{\kappa}\left(\ddot{\phi}+4H\dot{\phi}+\frac{1}{4}\dot{\phi}^2\right)e^{-\phi}.
\end{eqnarray}
Thus, the equation of state parameter (EOS) is given by
\begin{equation}\label{nva3}
\omega_{\phi}=-\left[1+\frac{\ddot{\phi}+H\dot{\phi}-\frac{1}{2}\dot{\phi}^2}{3\left(H\dot{\phi}+\frac{1}{4}\dot{\phi}^2\right)}\right].
\end{equation}
Using $\phi(t)=\phi_0(t_0/t)^n$, the expressions (\ref{nva1}), (\ref{nva2}) and (\ref{nva3}) become
\begin{eqnarray}\label{nva4}
\rho_{\phi}&=& \frac{3}{\kappa}\left(\frac{nH}{t}\phi-\frac{n^2}{4t^2}\phi^2\right)e^{-\phi},\\
\label{nva5}
p_{\phi}&=& \frac{1}{\kappa}\left[\left(\frac{n(n+1)}{t^2}-\frac{4nH}{t}\right)\phi+\frac{n^2}{4t^2}\phi^2\right],\\
\label{nva6}
\omega_{\phi}&=& -\left[1+\frac{\left(\frac{n(n+1)}{t^2}-\frac{nH}{t}\right)\phi-\frac{n^2}{2t^2}\phi^2}{3\left(-\frac{nH}{t}\phi+\frac{n^2}{4t^2}\phi^2\right)}\right].
\end{eqnarray}
It follows from (\ref{nva4}) that when $\phi<(4/n)(Ht)$ the energy density for the Weyl scalar field $\rho_{\phi}$ is positive. Moreover, for large times the linear term in $\phi$ inside the parentheses becomes dominant.\\

The condition (\ref{a19}) when we use (\ref{nva1}) and (\ref{nva2}) can be put in the form
\begin{equation}\label{nva8}
\omega_{\phi}<-\frac{1}{3}\left[1-\frac{\kappa(\rho_m+2\rho_r)e^{\phi}}{3(H\dot{\phi}+\frac{1}{4}\dot{\phi}^2)}\right]=-\frac{1}{3}\left(1+\frac{\Omega_m}{\Omega_{\phi}}+2\frac{\Omega_r}{\Omega_{\phi}}\right),
\end{equation}
where $\Omega_{\phi}=(8\pi G/3H_0^2)\rho_{\phi}$ is the density parameter associated to $\phi$.\\

According to the data combination Planck + WP + BAO + SN, in the present time $\Omega_{m_0}=0.307^{+0.011}_{-0.010}$, $\Omega_{tot_0}=1.000^{+0.0032}_{-0.0033}$ and $\Omega_{r_0}=2.47\cdot 10^{-5}h^{-2}$. Hence, using the expression $\Omega_{tot_0}=\Omega_{m_0}+\Omega_{r_0}+\Omega_{\phi_0}$, the condition (\ref{nva8}) nowadays becomes
\begin{equation}\label{nva9}
\omega_{\phi_0}<-0.481^{+0.0032}_{-0.0038}.
\end{equation}
Thus, the condition for accelerating expansion (\ref{a16}) is equivalent to the condition (\ref{nva9}) together with $\lbrace n>0, n<3p-1, p>1/3\rbrace$, for a power law Weyl scalar field. These requirements are not in contradiction with the data combination Planck + WP + BAO + SN, in which the present EOS parameter ranges in the interval $\omega_0=-1.10^{+0.08}_{-0.07}$ \cite{data}.\\

Now from (\ref{nva6}) it can easily be seen that for a fixed present EOS parameter $\omega_{\phi_o}$ the present value for the Weyl scalar field is restricted to the value
\begin{equation}\label{nva10}
\phi_0=\frac{4}{n(3\omega_{\phi_0}+1)}[H_0t_0(3\omega_{\phi_0}+4)-(n+1)].
\end{equation}
Using the formula $H_0t_0\simeq (2/3)\Omega_m$ we have the va\-lues $H_0t_0\simeq 0.2046^{+0.0074}_{-0.0066}$. Thus employing  that the EOS parameter  $\omega_{\phi_0}=-1.10^{+0.08}_{-0.07}$ (where we have identified $\omega_0$ with $\omega_{\phi_0}$) and taking the intermediate values for $H_0t_0$ and $\omega_{\phi_0}$ the expression (\ref{nva10}) reads
\begin{equation}\label{nva11}
\phi_0\simeq 1.7391-\frac{1.9881}{n}.
\end{equation}
The condition to the energy density (\ref{nva4}) to be positive in the present time, $\phi_0<(4/n)(H_0t_0)$, together with (\ref{nva11}) leads to the restriction $n<1.6137$. Therefore the condition $n<3p-1$ yields $p>0.8712$ which is compatible with the requirement $p>1$ for accelerating expansion models of the universe. Finally, it can be shown by direct substitution of (\ref{nva11}) that the deceleration parameter (\ref{a21}) becomes $q_0\simeq -1$ in the present time.\\

\section{Final Comments}

In this letter we have investigated cosmological solutions of a general theory of relativity formulated on a Weyl-integrable geometry, as it was introduced recently by C. Romero and collaborators in \cite{Carlos1,Carlos2}. We found the conditions under which the cosmological solutions exhibit acceleration in the scale factor. When these conditions hold, the Weyl scalar field, as the scalar part of the gravitational field, is capable to generate an acceleration in the expansion of the universe compatible with the observational data combination Planck+WP+BAO+SN, without the introduction of a dark energy component.\\

In this approach the action (\ref{a1}) respects the conformal equivalence principle and then, when a Palatini variational principle is applied to the action, the appropriate background geometry as a result is found to be the Weyl-integrable geometry \cite{Quiros}. The Riemann geometry is a particular case of the Weyl-integrable one, and in the Riemann frame the action (\ref{a1}) becomes the usual action for general relativity, where no Weyl scalar field appears and, of course, where to explain the accelerated expansion of the universe the introduction is necessary a dark energy component. Hence, we can interpret the Weyl scalar field as a gravitational field that in the Riemann frame is hidden. However, as was mentioned in \cite{Quiros}, both frames can be viewed as alternative descriptions of the same physical reality. The main difference is that in a Weyl frame the acceleration in the expansion is just a pure gravitational effect, whereas in the Riemann frame the dark energy component necessarily leads to the question about the origin and dynamics of this exotic component, which is a problem in the majority of the cosmological models.\\

\section*{Acknowledgements}

\noindent J.E.M.A and C.M. acknowledge CONACYT M\'exico and Universidad de Guadalajara  for financial support. R. A. acknowledges Universidad de Guadalajara for financial support. M.B. acknowledges CONICET and UNMdP for financial support.


\bigskip

\begin{thebibliography}{99}
\bibitem{Goenner} H.F.M. Goenner, Living Rev. Relativity {\bf 7}  (2004) 2.
\bibitem{Quint} D. Sapone, Int. J. Mod. Phys. {\bf A25} (2010), 5253-5331.
\bibitem{Carlos1} C. Romero, J. B. Fonseca-Neto, M. L. Pucheu, Found. Phys. {\bf 42} (2012) 224-240.
\bibitem{Carlos2} C. Romero, J. B. Fonseca-Neto, M. L. Pucheu, Class. Quant. Grav. {\bf 29} (2012) 155015.
\bibitem{BDcite1} C. Brans and R. H. Dicke, Phys. Rev. {\bf 124}, (1961) 925.
\bibitem{BDcite2} R. H. Dicke, {\it ibid}, {\bf 125}, (1962) 2163.
\bibitem{Romero1} T. S. Almeida, M. L. Pucheu, C. Romero and J. B. Formiga, Preprint arXiv:1311.5459 (2013).
\bibitem{Quiros} I. Quiros, R. Garcia-Salcedo, J. E. Madriz-Aguilar and T. Matos, Gen. Relativ. Gravit. {\bf 45} (2013) 489-518.
\bibitem{data} J. Beringer et al. (Particle Data Group), Phys. Rev. {\bf D86}, (2012) 010001.
\bibitem{Cheng} H. Cheng, Phys. Rev. Lett. {\bf 61} (1998) 2182, ArXiv:math-ph/0407010.

\end{thebibliography}
\end{document}